\documentclass{article}
\usepackage{amsmath, amssymb}

\renewcommand{\abstractname}{Abstract.}
\renewcommand\abstract{\hfil\break\topsep=0pt\partopsep=0pt\parsep=0pt\itemsep=0pt\relax
\trivlist\item[\hskip\labelsep {\bfseries\abstractname}]\if!\abstractname!\hskip-\labelsep\fi}

\newcommand{\email}[1]{{(e-mail: #1)}}

\def\keywordname{{\bfseries Key words:}}
\def\keywords#1{\par\addvspace\baselineskip\noindent\keywordname\enspace
\ignorespaces#1}

\def\jelclassname{{{\bfseries JEL Classification:}} }
\def\jelclass#1{\par\addvspace\medskipamount\noindent\jelclassname\
\ignorespaces#1}

\def\title#1{\hfil\break\hfil\break
\hfil\break\par\addvspace\baselineskip\noindent \ignorespaces{\LARGE\bf#1}\hfil\break}

\def\author#1{\par\addvspace\baselineskip\noindent
\ignorespaces{\large\bf#1}}

\def\institute#1{\par\addvspace\baselineskip\noindent
\ignorespaces{\small#1}\hfil\break}

\usepackage{enumerate}
\usepackage{graphicx}

\numberwithin{equation}{section} \overfullrule=0pt

\newcommand{\eq}[1]{(\ref{#1})}

\newcommand{\UqX}{U^q_X}

\newcommand{\barX}{{\bar X}}
\newcommand{\uX}{{\underline X}}

\newcommand{\UqbX}{U^q_{\barX}}
\newcommand{\UquX}{U^q_{\uX}}

\newcommand{\bfo}{{\bf 1}}

\newcommand{\bbe}{\begin{equation}}

\newcommand{\INT}{\int_{-\infty}^{+\infty}}

\newcommand{\cK}{{\mathcal K}}

\newcommand{\R}{{\bf R}}

\newcommand{\be}{\beta}
\newcommand{\bep}{\beta^+}
\newcommand{\bem}{\beta^-}

\newcommand{\bepo}{\be^+_1}
\newcommand{\bept}{\be^+_2}

\newcommand{\ap}{a^+}
\newcommand{\am}{a^-}
\newcommand{\apo}{a^+_1}

\newcommand{\apt}{a^+_2}

\newcommand{\cp}{c^+}
\newcommand{\cm}{c^-}

\newcommand{\cpo}{c^+_1}
\newcommand{\cpt}{c^+_2}

\newcommand{\ee}{\end{equation}}

\newcommand{\la}{\lambda}
\newcommand{\lp}{\lambda^+}
\newcommand{\lm}{\lambda^-}
\newcommand{\sgp}{\sigma^+}
\newcommand{\sgm}{\sigma^-}
\newcommand{\lpo}{\lambda^+_1}
\newcommand{\lpt}{\lambda^+_2}

\newcommand{\kapq}{\kappa^+_q}
\newcommand{\kamq}{\kappa^-_q}

\newcommand{\Om}{\Omega}

\newcommand{\ga}{\gamma}

\newcommand{\ka}{\kappa}
\newcommand{\sg}{\sigma}

\begin{document}
\title{Practical guide to real options in discrete time}
\author{Svetlana Boyarchenko
$^{\ast}$ }\footnote{Corresponding author}
\author{Sergei Levendorski\v{i}$^\dagger$}

\institute{ $^{\ast}$ Department of Economics, The University of Texas at Austin, 1 University Station C3100,
Austin, TX 78712-0301, U.S.A. \email{sboyarch@eco.utexas.edu}\\ $^\dagger$ Department of Economics, The University
of Texas at Austin, 1 University Station C3100, Austin, TX 78712-0301, U.S.A. \email{leven@eco.utexas.edu}}


\begin{abstract}Continuous time models in the  theory
of real options give explicit formulas for optimal exercise strategies when options are simple and the price of an
underlying asset follows a geometric Brownian motion. This paper suggests a general, computationally simple
approach to real options in discrete time. Explicit formulas are derived even for embedded options. Discrete time
processes reflect the scarcity of observations in the data, and may account for fat tails and skewness of
probability distributions of commodity prices. The method of the paper is based on the use of the expected present
value operators. 
\end{abstract}


\jelclass{D81, C61, G31}


\section{Introduction}
The goal of the paper is to provide a general framework for
pricing of real options in discrete time; since option pricing
cannot be separated from the problem of the optimal exercise time
of an option, we solve for the optimal timing as well. Valuation
of real options and optimal exercise strategies explicitly derived
in the paper are relevant to many practical situations where
individuals make at least partially irreversible decisions. Simple
real options cover such situations as timing an investment of a
fixed size,  the scrapping of a production unit,  a capital
expansion program, etc. Embedded real options are relevant in such
cases as new technology adoption, timing  an investment
(partially) financed by debt with an embedded option to default in
the future, human capital acquisition, and many others (see, for
example, \cite{DP},  papers in the volume \cite{BrT}, and the
bibliography therein). Usually, continuous time models are used.
We believe that the discrete time approach has certain advantages
over the continuous time approach to modeling of real options.

The discrete time approach allows one to incorporate in a more
tractable way such important aspects of economic reality as
``time-to-build". Some options -- for example, renewing a labor
contract or awarding a grant to a research project -- can be
exercised only at certain dates (typically, once a year). Further,
the data in economics is much more scarce than in finance, where
continuous time models were originated and developed.
Observations of some variables are available only quarterly or
even annually, and
it is easier to fit a transition density to the data in a discrete time model than in a continuous time one,  when
only a  list of observations at equally spaced moments in time is available. There are also  natural questions,
such as the rates of job creation and job destruction (see, e.g., \cite{DHS} and \cite{CQ}), which cannot be
addressed in the Brownian motion setting under the standard assumption that labor is instantly adjustable, because
the rates will be infinite. In the discrete time model, one obtains finite closed form solutions.
Finally, discrete time modeling is natural for economics, where
agents do not make, say, investment decisions every instant unlike
the investors in financial markets.

In fact, the appearance of  continuous time models in economics was mainly due to the success of these models in
finance, and their tractability. Notice, however, that even in finance,  the
success of simple models is  gone.
Relatively simple models use the (geometric) Brownian motion as
the underlying stochastic process (see, for example, \cite{Du}).
However, these models proved to be rather inaccurate, and have
been amended in many ways, none of which being as tractable as the
Brownian motion model. In particular, nowadays processes with
jumps are widely used (see, for instance, \cite{CaW} and
\cite{DPS}). The normality of commodity price processes is
rejected by the data as well (see, for example, \cite{DL} or
\cite{YB}). Assume, for the moment, that we are willing to
sacrifice the accuracy of a model for its simplicity, and decide
to use the Brownian motion model. The apparent advantage is a
well-known manageable scheme:  with the help of Ito's lemma write
down a second order differential equation  for the value of an
option, employ economic arguments to add appropriate boundary
conditions, such as value matching and smooth pasting, and, using
the general solution to the differential equation, reformulate the
problem as a system of algebraic equations. In elementary
situations, a closed form solution can be derived, and a simple
exercise strategy results; in other situations, numerical
procedures are available.

Unfortunately, this scheme is  uncomplicated only when one
considers really simple options. If a part of the option value
comes from instantaneous payoffs due at certain future dates, a
closed form solution to the optimal exercise problem is no longer
available. If we consider an option whose value comes from
different streams in different regions of the state space, the
resulting system of algebraic equations may involve too many
unknowns, and it becomes messy indeed. If we consider embedded
options, then there is no general result about the optimal
exercise rule, apart from the heuristic smooth pasting condition,
and it is not even clear whether the formal solution satisfying
the smooth pasting condition exists. In addition, if one
incorporates jumps, the intuitive justification  for the smooth
pasting condition in \cite{DP} is lost, and there is no reason to
believe that this principle always holds\footnote{We are grateful
to Avinash Dixit for pointing out to us that this principle may
not hold in a discrete time model.}. In fact, it may fail as it
was shown in \cite{BL3} and \cite{BL4}.

From our point of view, the following two drawbacks of the standard approach are conceptually even more important.
First, the formulas provided by the standard approach are certain analytic expressions, which   have no clear
economic meaning by themselves, and the optimal exercise boundaries are just numbers. The classical example is the
investment rule formulated in \cite{DP} in terms of the correction factor to the Marshallian law: the correction
factor is  an expression in terms of a root of a certain quadratic polynomial (fundamental quadratic).
Second, is it really necessary to write down the equation for the value function, the boundary conditions, and
solve the (free) boundary  problem? Indeed, the option value by definition is the expected present value (EPV) of
an instantaneous  payoff or stream of payoffs. Suppose, we have formulas for these expected present values for
model payoffs. Then  using the argument similar to the backward induction in discrete time models with the finite
time horizon, we can calculate the option value step by step. We solve for the value and optimal exercise time (if
necessary) of the most distant option in a sequence of embedded options, then we move to the second to last
option, etc.

This paper provides a general framework of this sort, with
explicit formulas for each model situation, under the assumption
that  the underlying stochastic process is a random walk (under a
risk-neutral measure chosen by the market), $X_t$, on the real
line, that is, $X_t=X_0+Y_1+Y_2+\cdots +Y_t$, where $Y_1, Y_2,
\ldots $ are independently identically distributed random
variables on a probability space $\Om$, and $X_0$ is independent
of $Y_1, Y_2, \ldots$ This process specification implies that the
dates when observations and/or decisions to exercise  options can
be made are equally spaced, and time periods are normalized to
one.
 We calculate the EPV's in the following
model situations: (1) an instantaneous payoff tomorrow; (2) a perpetual stream of payoffs; (3) an instantaneous
payoff which is due when a barrier is crossed from below (or from above); (4) an entitlement for a stream of
payoffs which comes into effect when a barrier is crossed from below (or above); (5) a stream of payoffs which
will be lost when a barrier is crossed from below (or from above). Using these EPV's, we solve for the  optimal
exercise time of an option, calculate the expected waiting time till an option will be exercised, and find the
optimal gradual capital expansion strategy.

 Let $g(X_t)$ be the payoff function, and $q\in
(0,1)$ the discount factor. The first two cases are simple: the
calculation of $qE^x[g(X_1)]$, and the summation of an infinite
series $\sum_{t=0}^{\infty} q^t E^x[g(X_t)]$, respectively.  For
$g(X_t)$ a linear combination of exponents, the calculations in
the first two cases reduce to the calculation of values of the
moment generating function of $Y_1$, and closed-form solutions
result. These observations are fairly standard and commonly used.
The novelty of our approach is that the solutions in all the cases
above are expressed in terms of the expected present value
operators (EPV--operators).  For a random walk $X$, and a given
stream of payoffs $g(X_t)$, the EPV--operator (denoted by $\UqX$)
calculates the EPV of the stream $g(X_t)$:
\[ \UqX g(x)=E^x\left[\sum_{t=0}^\infty q^t
g(X_t)\right]=E\left[\sum_{t=0}^\infty q^t g(X_t)\ |\
X_0=x\right].
\]
This operator gives the EPV in  case (2). Cases (3) and (5) reduce
to cases (2) and (4). To express the EPV of a payoff in case (4),
 in addition
to the EPV--operator of $X$, we need the EPV--operators of the
supremum process $\barX_t=\max_{0\le s\le t}X_s$ and the infimum
process $\uX=\min_{0\le s\le t}X_s$. These EPV--operators
 act as follows
 \[
 \UqbX g(x)=E^x\left[\sum_{t=0}^\infty q^t
g(\barX_t)\right]\equiv E\left[\sum_{t=0}^\infty q^t g(\barX_t)\
|\ X_0=x\right],
 \]
\[
 \UquX g(x)=E^x\left[\sum_{t=0}^\infty q^t
g(\uX_t)\right]\equiv E\left[\sum_{t=0}^\infty q^t g(\uX_t)\ |\
 X_0=x\right].
 \]
 Let $\bfo_{[h, +\infty)}$ denote the indicator function of the interval $[h,
 +\infty)$ and the multiplication operator by the same function.
 We prove that the EPV of a stream $g(X_t)$  that accrues after
 the barrier $h$ is crossed from below is the composition of the
 EPV--operators $\UqbX$ and $\UquX$, and $\bfo_{[h, +\infty)}$:
 \begin{equation}\label{bar1}
 V(h; x)= (1-q)\UqbX \bfo_{[h, +\infty)}\UquX g(x).
 \end{equation}
 Similar formulas are obtained for the other expected present values listed in
 cases (3)--(5).

 Assume that both the supremum and infimum processes are non-trivial, i.e., jumps in both
 directions happen with non-zero probabilities. Then the form of the solution \eq{bar1} and simple general properties
 of the EPV--operators allow us to give a very short (approximately a
 page) proof of the optimal exercise rule: exercise an option which yields a stream of
 increasing payoffs
 $g(X_t)$ the first time when the EPV of the
 stream $g$ under the {\em infimum process} becomes non-negative
 \footnote{For instance, in timing an
 investment of a fixed size, $g$ is the stream of revenues net of
 $(1-q)C$, where $C$ is the fixed cost of investment.}.
 In this case, the {\em record
 setting low} payoff process matters. This reflects Bernanke's {\em bad news principle} spelled out in \cite{Ber}.
Similarly, for  a decreasing payoff function $g$,
 the {\em record setting high} payoff process matters: exercise the option the first time
 when the EPV of the stream $g$ under the {\em supremum process} becomes non-negative.
 Notice that a sufficient condition for  optimality makes a perfect
 economic sense: the payoff function $g(X_t)$ is a monotone function of the
 stochastic factor $X_t$.  {\em Record setting news
 principles} were obtained in \cite{BO} in the continuous
 time model, for a special case of payoff streams of the form
 $Ge^x-C(1-q)$ or $C(1-q)-Ge^x$, by using the well-known results for the
 perpetual American call and put options, respectively. The present paper
 extends the record setting news principles to  general continuous monotone
 streams in  discrete time models. If the payoff stream, $g$, is an increasing function of the stochastic factor
 $X_t$, then it is optimal to exercise the right for  (respectively, to give up) the stream when the EPV of
 the stream calculated under the
 assumption that the process $\{X_t\}$ is replaced by the infimum process becomes non-negative (respectively,
 non-positive). For a decreasing payoff function $g(X_t)$, one only needs to substitute the supremum process for the
 infimum process in the above statement.

The reader may think that the calculation of the action of the expected present value operators is difficult. For
a general random walk, this is really the case. However, one of the advantages of the discrete time setting is
that the transition density of a random walk can be approximated by exponential polynomials on each half-axis with
desired accuracy and simplicity, although there is certainly a trade-off between the two. The family of transition
densities given by exponential polynomials is fairy flexible, and such densities can account for fat tails and
skewness observed in empirical distributions of commodity prices. For a transition density of this family, and
$g(x)$ given by exponential polynomials (different exponential polynomials on different intervals of the
 real line are admissible), all the functions $\UqX g(x)$, $\UqbX g(x)$ and
 $\UquX g(x)$ are of the same class as $g(x)$. Therefore, we can use
 general formulas derived for model situations several times (embedded options),
 and each time we will calculate the integrals
 of a simple structure. All the  tools needed for the
  calculation of these integrals are the fundamental theorem of
  calculus
  and integration by parts. It is unnecessary to write down
  differential equations and solve them. In fact, after the roots
  of a certain polynomial are found (we will call it the {\em fundamental polynomial}
  by analogy with the {\em fundamental quadratic} in the Brownian motion
  case: see \cite{DP}), all the calculations reduce to
  straightforward algebraic manipulations. Even in the case of
  embedded options of complicated structure, there is no need to consider
  systems of algebraic equations, and  timing an option reduces to
  the calculation of a (unique) zero of a monotone function. At the same time, the method of
  the paper allows one to solve the optimal exercise problems for
  much wider classes of payoffs than the standard approach
  does.

  The rest of the paper is organized as follows. In Section 2, we
  present general formulas for EPV's in model situations (1)--(5)
  and determine the optimal exercise time and the waiting time
  till an option is exercised. In Section 3, explicit formulas for
  the results of Section 2 are presented. In Section 4, timing a
  capital expansion program is solved by reduction to a sequence
  of investment opportunities of fixed size, and formulas
  for the optimal capital stock and the value of the firm are
  derived. The last formula shows clearly that the value of the firm does not evolve as a Markovian
  process. A similar result can be obtained in the continuous time
  model.
  Therefore the value of a firm cannot be modeled as a Gaussian process as it is often done in finance.
  Technical details and proofs are given in the appendices.

\section{Model situations}
\subsection{Instantaneous payoff and payoff stream}
The standing assumption of the paper is that the underlying
stochastic process $X_t$ is a random walk on $\R$,
which has the transition density, $p$ (the method of the paper can
be applied to random walks on lattices as well). The transition
operator, $T$, is defined by
\[(Tf)(x)=E^x[f(X_1)]\equiv E[f(X_1)\vert\,X_0=x].\]
Given $p$,
one calculates  the EPV of a stochastic
payoff tomorrow:
\[E^x[qg(X_1)]=q(Tg)(x)=q\INT p(y)f(x+y)dy.\]
To compute the EPV of a stochastic payoff $t$ periods from now, we
use the Markov property of a random walk:
\[E^x[q^tg(X_t)]=q^t(T^tg)(x).\]
The next step is to calculate the EPV of a stream of payoffs:
\[\UqX g(x) =E^x\left[\sum_{t=0}^\infty q^tg(X_t)\right]=\sum_{t=0}^\infty q^t(T^tg)(x).\]
Under the condition $q\in(0,1)$, the operator $I-qT$ is invertible in $L^\infty(\R)$, and the inverse is
$(I-qT)^{-1}=\sum_{t=0}^\infty q^tT^t$, hence
\begin{equation}\label{uqx}
\UqX g(x)=(I-qT)^{-1}g(x),
\end{equation}
for a bounded measurable $g$. Under additional conditions on the
transition density, \eq{uqx} can be extended to unbounded $g$.
From \eq{uqx}, we conclude that
\begin{equation}\label{trans}
\UqX(I-qT)=(I-qT)\UqX=I.
\end{equation} This equation allows one to
express an instantaneous payoff $G(X_t)$ in terms of the EPV of a
stream and vice versa:
\begin{eqnarray}\label{inst} g(X_t)&=&(I-qT)G(X_t),\\
\label{stream} G(X_t)&=&\UqX g(X_t).\end{eqnarray} Clearly, equation \eq{stream} is the value, at date $t$, of the
stream of payoffs that starts at date $t$ as well. To evaluate such a stream at the initial date, we compute
\bbe\label{stream0} E^x[q^tG(X_t)]=q^tT^t\UqX g(x).\ee

\subsection{Payoff stream that starts to accrue at a random time}
The list of the EPV's presented in the previous subsection conveys no new information to any person familiar with
the basics of economics or finance. Suppose that we want to price the stream of payoffs that starts to accrue
after the underlying stochastic variable, $X_t$, crosses a certain barrier, $h$. In this case, we need to replace
the deterministic time, $t$, on the LHS in \eq{stream0} with a random time, $\tau$, and \eq{stream0} is no longer
valid. Fortunately, it is still possible to compute the value $E^x[q^{\tau}f(X_{\tau})]$ in terms of the EPV's of
some payoff streams. To do this, we need to distinguish between two cases: $X_t$ crosses $h$ from below, and $X_t$
crosses $h$ from above. Denote $\tau^+=\tau^+_h=\min\{t\ | \ X_t\ge h\}$, and consider
\[V^+(x;h)=E^x\left[\sum_{t=\tau^+}^\infty q^tg(X_t)\right].\]
Under  natural conditions on  $g$,
\begin{equation}\label{whp}
V^+(x;h)=E^x\left[q^{\tau^+} \UqX
 g(X_{\tau^+})\right]=(1-q)\UqbX \bfo_{[h, +\infty)} \UquX g(x).
 \end{equation}
 Equation \eq{whp} holds due to the quite deep result, known as the
 Wiener-Hopf factorization formula, that establishes the
 relationship among the EPV--operators $\UqX,\, \UqbX,$ and
 $\UquX$:
\begin{equation}\label{wh}
 \UqX
 =(1-q)\UqbX \UquX
 \end{equation}
 (for details, see Subsection 3.2).
 The Wiener-Hopf factorization formula \eq{wh} per se has no
 economic intuition behind it (and neither does Ito's lemma, both being general deep mathematical
 results), but using
 this formula, one obtains values of real options and  optimal
 exercise strategies in economically meaningful terms.

 Similarly, for the case of crossing from above, set
 $\tau^-=\tau^-_h=\min\{t\ | \ X_t\le h\}$ and
\[V^-(x;h)=E^x\left[\sum_{t=\tau^-}^\infty q^tg(X_t)\right].\]
Then the value of such a stream is given by
\begin{equation}\label{whm}
V^-(x;h)=E^x\left[q^{\tau^-} \UqX g(X_{\tau^-})\right]=(1-q)\UquX
\bfo_{(-\infty, h]}\UqbX g(x).
\end{equation}
It is easy to see that \eq{whm} is just a version of \eq{whp}:
change the direction on the real line, then the supremum process
becomes the infimum process, and vice versa.

Notice that \eq{whp} and \eq{whm} price the streams of payoffs
that are described in case (4). The model situations in case (3)
reduce to these formulas if we express an instantaneous payoff as
the EPV of a stream using \eq{inst}:
\begin{eqnarray*}E^x\left[q^{\tau^+}
 G(X_{\tau^+})\right]&=&(1-q)\UqbX \bfo_{[h, +\infty)} \UquX (I-qT)G(x);\\
E^x\left[q^{\tau^-}
 G(X_{\tau^-})\right]&=&(1-q)\UquX \bfo_{(-\infty, h]} \UqbX (I-qT)G(x).
 \end{eqnarray*}
\subsection{Payoff stream that is lost at a random time}
The case when the right to the stream $g(X_t)$ is lost when $X_t$
crosses $h$ from below is reduced to the cases considered above:
\begin{eqnarray*}V_+(x;h)&=&E^x
\left[\sum_{t=0}^{\tau^+-1}q^tg(X_t)\right]\\
&=&E^x\left[\sum_{t=0}^{\infty}q^tg(X_t)\right]-E^x\left[\sum_{t=\tau^+}^{\infty}q^tg(X_t)\right]\\
&=&\UqX g(x)-(1-q)\UqbX \bfo_{[h, +\infty)} \UquX g(x).
\end{eqnarray*}
The Wiener-Hopf factorization formula \eq{wh} allows us to
proceed: \begin{eqnarray}\nonumber
V_+(x;h)&=&(1-q)\UqbX\left(\UquX g(x)-\bfo_{[h, +\infty)} \UquX
g(x)\right)\\ \label{whpp} &=&(1-q)\UqbX \bfo_{(-\infty,h)} \UquX
g(x).\end{eqnarray}

In a similar way, if the right for the stream is lost on crossing
$h$ from above, we derive \bbe\label{whmm}
V_-(x;h)=E^x\left[\sum_{t=0}^{\tau^--1}q^tg(X_t)\right]=(1-q)\UquX
\bfo_{(h, +\infty)} \UqbX g(x).\ee 
\subsection{Optimal exercise time}
\subsubsection{Time to enter}
Suppose now that $g$ is the stream of payoffs specified for a real
option, for example, an option to invest capital $C$ into a
technology. If an entrepreneur chooses to invest at time $\tau$,
then the technology will produce a commodity at rate $G$, starting
from time $\tau+1$, ever afterward. Investment is irreversible.
One may view $C$ as the present value of a deterministic stream of
expenditures, to which the investor commits at time $\tau$. By
\eq{inst}, this stream is $c=(1-q)C$. The output is sold on the
spot at the market price $P_t=e^{X_t}$. Obviously, we are facing
the situation to which \eq{whp} applies with
$g(X_t)=qGE_t[e^{X_{t+1}}]-(1-q)C$. In Appendix A, we show that if
$g$ is non-decreasing and $\UquX g(x)$ changes sign only once, the
optimal exercise (log) price (the optimal investment threshold in
our example), $h^*$ is a unique solution to the equation
\begin{equation}\label{thrp1}
w(x)=(\UquX g)(x)=0.
\end{equation}
We stress that contrary to the Brownian motion  model and other models with continuous trajectories, the exercise
boundary in the present model is not necessarily the price at which the option is exercised. It is optimal to
exercise the option the first time $\tau$ such that $X_\tau\ge h^*$.
For the case of irreversible investment, \eq{thrp1} is equivalent
to \bbe\label{thrp2} E^x\left[\sum_{t=0}^\infty
q^t\left(qGe^{\uX_{t+1}}-(1-q)C\right)\right]=0.\ee The last
equation says that investment becomes optimal when the EPV of the
project calculated under the assumption that the original price
process is replaced by the infimum process becomes non-negative.
If the investment threshold is chosen optimally, then the value of
the option to invest, as given by \eq{whp}, is
\[
V^+(x;h^*)=(1-q)\UqbX \bfo_{[h^*, +\infty)} \UquX g(x).
 \]
 We use \eq{thrp2} to write
 \bbe\label{cost}
 C=e^{h^*}E\left[\sum_{t=1}^\infty q^tGe^{\uX_t}\,\vert\,X_0=0\right].\ee
Therefore, \begin{eqnarray*}E^x\left[\sum_{t=0}^\infty
q^t\left(qGe^{\uX_{t+1}}-(1-q)C\right)\right]&=&e^xE\left[\sum_{t=1}^\infty
q^tGe^{\uX_t}\,\vert\,X_0=0\right]-C\\&=&(e^x-e^{h^*})Ce^{-h^*},\end{eqnarray*}
and
\begin{eqnarray*}
V^+(x;h^*)&=&Ce^{-h^*}(1-q)(\UqbX \bfo_{[h^*,
+\infty)}(\cdot)(e^\cdot-e^{h^*}))(x)\\
&=&Ce^{-h^*}(1-q)E^x\left[\sum_{t=0}^\infty
q^t(e^{\barX_t}-e^{h^*})_+\right],\end{eqnarray*} where
$(e^{\barX_t}-e^{h^*})_+=\max\{(e^{\barX_t}-e^{h^*}),0\}$. Now we
can express Tobin's $Q$  as
\[Q(x)=\frac{V^+(x;h^*)}{C}=\frac{E^x\left[\sum_{t=0}^\infty
q^t(e^{\barX_t}-e^{h^*})_+\right]}{e^{h^*}/(1-q)}.\] Hence Tobin's
$Q$ is the ratio of the EPV of the stream of payoffs
$(e^{\barX_t}-e^{h^*})_+$ and the EPV of the perpetual stream
$e^{h^*}$. Notice that the investment threshold is determined by
the infimum process, and  Tobin's $Q$ is determined by the
threshold and supremum process.

Alternatively, using \eq{cost}, we can write the value
$V^+(x;h^*)$ as
\[V^+(x;h^*)=(1-q)E\left[\sum_{t=1}^\infty q^tGe^{\uX_t}\,\vert\,X_0=0\right]E^x\left[\sum_{t=0}^\infty
q^t(e^{\barX_t}-e^{h^*})_+\right].\] The last representation factors out contributions of the infimum and supremum
price processes to the value of the option to invest. The first expectation on the RHS decreases as the
probability of downward jumps in prices increases, and the second expectation increases with the probability of
upward jumps. If both probabilities increase, the overall effect on the value $V^+(x;h^*)$ is ambiguous.
\subsubsection{Time to exit}
The option considered above is similar to the perpetual American
call option. In this subsection, we consider the optimal exercise
strategy for a generalization of the perpetual American   put
option, which is relevant to irreversible decisions such as
scrapping a production unit or exit; it also applies to the
optimal timing of default on a debt.

Consider a firm in a deteriorating environment. The firm's profit,
$f(X_t)$, is falling, on average, so that  it may become optimal
to discontinue operations at some point in time, and sell  the
firm's inventory for the scrap value $C$. If the firm's manager
makes a decision to scrap the inventory at time $\tau$, the
operations will be discontinued at time $\tau+1$, and the scrap
value will be available at time $\tau+2$. When the firm
discontinues its operations, it looses the stream of profits whose
EPV is \[E^x\left[\sum_{t=\tau+1}^\infty
q^tf(X_t)\right]=E^x[q^{\tau+1}\UqX f(X_{\tau+1})].\] The scrap
value, $C$, can be viewed as a deterministic stream of payoffs,
$c=(1-q)C$, that starts to accrue at $\tau+2$.  Hence, at time
$\tau$, the option value of exit is $\UqX g(X_\tau)$, where
$g(x)=q^2(1-q)C-qTf(x)$. If the decision to exit is made when the
underlying stochastic factor, $X_t$, crosses a barrier $h$ from
above, then the option value to exit is given by \eq{whm}. Suppose
that $g(x)$ is non-increasing and $\UqbX g(x)$ changes sign only
once. Then the optimal exercise price of the option is a unique
solution, call it $h_*$, of the equation
\begin{equation}\label{thrscr}
w(x)=(\UqbX g)(x)=0
\end{equation}
(see Appendix A). For $g$ in the case under consideration,
\eq{thrscr} is equivalent to \bbe\label{exit1}
E^{x}\left[\sum_{t=0}^\infty
q^t(q^2(1-q)C-qf(\barX_{t+1}))\right]=0,\ee or \bbe \label{exit}
f(x)+q^2C=\UqbX f(x). \ee The LHS in \eq{exit} is the EPV of the
instantaneous payoff received if the operations are terminated,
the RHS is the EPV of the stream of profits evaluated under the
assumption that the original stochastic process is replaced by the
supremum process. Rule \eq{exit} says that it is optimal to exit
when the difference between the former and the latter EPV's
becomes non-negative.

 Suppose that $f(x)=Ge^x$, then
from \eq{exit1}, one obtains
\[q^2C=e^{h_*}E\left[\sum_{t=1}^\infty q^tGe^{\barX_t}\,\vert\,X_0=0\right],\]
whence \[E^x\left[\sum_{t=0}^\infty
q^t(q^2(1-q)C-qf(\barX_{t+1}))\right]=q^2Ce^{-h_*}(e^{h_*}-e^x),\]
and from \eq{whm}, we derive
\begin{eqnarray*}V^-(x;h_*)&=&q^2Ce^{-h_*}(1-q)\left(\UquX
\bfo_{(-\infty, h_*]}(\cdot)(e^{h_*}-e^\cdot)\right)(x)\\
&=&q^2Ce^{-h_*}(1-q)E^x\left[\sum_{t=0}^\infty
q^t(e^{h_*}-e^{\uX_t})_+\right]\\
&=&(1-q)E\left[\sum_{t=1}^\infty
q^tGe^{\barX_t}\,\vert\,X_0=0\right]E^x\left[\sum_{t=0}^\infty
q^t(e^{h_*}-e^{\uX_t})_+\right].
\end{eqnarray*}

\subsection{Expected waiting time} We consider timing the investment of a fixed size
once  again. Assume that the spot log-price $x$ is less than
$h^\ast$, and consider the waiting time $R_x$ till the investment
is made. This is the random variable defined by
\[
R_x=\min\{t>0\ |\ X_t\ge h^\ast\}.
\]
The expected waiting time can be calculated as follows:
\begin{eqnarray}\nonumber
E[R_x] &=&E^x\left[\sum_{t=0}^\infty {\bf 1}_{(-\infty,
h^\ast)}(\barX_t)\right]=\lim_{q\to 1-0}E^x\left[\sum_{t=0}^\infty
q^t {\bf 1}_{(-\infty, h^\ast)}(\barX_t)\right]\\\nonumber
&=&\lim_{q\to 1-0}\left\{E^x\left[\sum_{t=0}^\infty q^t
\right]-\lim_{q\to 1-0}E^x\left[\sum_{t=0}^\infty q^t {\bf
1}_{[h^\ast, +\infty)}(\barX_t)\right]\right\}\\\label{wait0}
&=&\lim_{q\to 1-0}\left\{(1-q)^{-1}-\UqbX \bfo_{[h^\ast,
+\infty)}(x)\right\}.
\end{eqnarray}
Notice that here one must consider the process under the historical measure and not a risk-neutral one. Explicit
formulas for the expected waiting time are presented in Appendix C.

\section{Explicit formulas for model situations}

In this section, we show how to calculate the EPV when the
transition density is given by exponential polynomials. 
\subsection{Instantaneous payoff and payoff stream}

\subsubsection{Calculation of the expected present value of an
instantaneous payoff at a deterministic moment in the future} Let
$p$ be the transition density of the random walk
$X_t=X_0+Y_1+\cdots +Y_t$, and let
\[
M(z)=E\left[e^{zY_1}\right]=\INT e^{zx}p(x)dx \]
 be the moment
generating function of $Y_1$. If $G$ is an exponential function: $G(x)=e^{z x}$, then the transition operator,
$T$, acts as a multiplication operator by the number $M(z)$, and we obtain $qE^x[e^{zX_1}]=qT G(x)=q M(z)e^{zx}$.
Similarly,  one easily calculates the expected present value of a payoff $G(X_t)$ at a deterministic moment $t$ in
the future: $q^tE^x[e^{zX_t}]=((qT)^t G)(x)=(qM(z))^t e^{zx}$.
 Equations
\eq{inst} and \eq{stream} can be used to represent a payoff $G$ as the expected present value of the stream
$g(X_t)=((I-qT)G)(X_t)$; $G(X_t)=\UqX g(X_t)$.  In a special case of an exponential function $G(x)=e^{z x}$, with
$z$ satisfying the condition $1-qM(z)>0$, we have $g(x)=(1-qM(z))G(x)$.

\subsubsection{Expected present value of an ongoing project:
capital and labor are fixed} Consider a firm that produces a commodity which is sold on the spot at the price
$P_t=e^{X_t}$. We impose the following restriction on the price process: \bbe\label{bd2} q M(1) <1.
\end{equation}
Restriction \eq{bd2} is appropriate when  capital, $K$, and labor, $L$, employed by the firm are fixed or capped.
Indeed, for $K$ and $L$ fixed, the EPV of the revenue stream grows each period by  factor
$M(1)=E\left[e^{Y_1}\right]$, and it is discounted back at  rate $q$. Hence, the EPV of the stream of revenues is
given by
\begin{equation}\label{epv1}
P_0G(K, L)\sum_{j=0}^\infty (qM(1))^t=\frac{P_0G(K, L)}{1-qM(1)},
\end{equation}
where $G(K, L)$ is the production function of the firm. For the series in \eq{epv1} to converge, it is necessary
and sufficient that \eq{bd2} holds.

\subsubsection{The case of costlessly adjustable labor} In other
situations, more stringent conditions than \eq{bd2} may be necessary. For instance, consider a firm with the
Cobb-Douglas production function $G(K, L)=dK^\theta L^{1-\theta}$, which faces the fixed labor cost $w$, and can
instantly and costlessly adjust labor. A similar situation was considered in \cite{AbE} for a continuous time
model of irreversible investment. At a given price level of the output, $P$, the firm chooses $L$ to maximize
$PdK^\theta L^{1-\theta}-wL$. From the F.O.C.
\[
(1-\theta)PdK^\theta L^{-\theta}-w=0,
\] we find
$ L=K\bigl((1-\theta)Pd/w\bigr)^{1/\theta}, $ and hence the
revenue is $R=AP^{1/\theta}$, where
\[
A=Kw\left(\frac{(1-\theta)d}{w}\right)^{1/\theta}\cdot\frac{\theta}{1-\theta}.
\]
We conclude that the EPV of the revenue stream
\begin{equation}\label{epv2}
AP_0\sum_{t=0}^\infty (qM(1/\theta))^t=
\frac{AP_0}{1-qM(1/\theta)}
\end{equation}
 is finite iff
$qM(1/\theta)<1.$
 Notice that if  increases in prices are anticipated, that is,
 $M(1)>1$, then $M(1/\theta)>M(1)$ (recall that the moment
 generating function is convex), and therefore, the condition $qM(1/\theta)<1$ is more
 stringent than \eq{bd2}.

\subsubsection{The case of stochastic prices and operational costs}
One can allow for a  stochastic operational cost as well, and the
model remains quite tractable provided both prices and variable
costs depend on the same stochastic factor, call it $X_t$. A
simple example (with fixed capital and labor) would be
$R(X_t)=e^{X_t}G(K, L),$ and  $C(X_t)=a+be^{\ga X_t},$
where $a, b, \ga>0$, so that the profit is
\[
g(X_t)=R(X_t)-C(X_t)=e^{X_t}G(K, L)-a-be^{\ga X_t}.
\]
These examples show that it may be necessary to consider payoffs
of various kinds; hence,  some general regularity conditions on
$g$ are needed.

\subsubsection{Calculation of $\UqX g(X)$ for general payoffs}
The EPV of a stream $g(X_t)$ can be easily calculated when $g$ is
an exponential polynomial, that is, a sum of products of exponents
and polynomials. For $g(X_t)=e^{z X_t}$,
the result is
\begin{equation}\label{epv4}
\UqX e^{z x}=E^x\left[\sum_{t=0}^\infty q^t
g(X_t)\right]=\frac{e^{z x}}{1-qM(z)},
\end{equation}
provided $1-qM(z)>0$. By linearity, \eq{epv4} extends to $g$ a linear combination of exponential functions, and
more generally, for integrals of exponential functions. Hence, if an explicit formula for the moment generating
function is available,  $\UqX g(X)$ can be calculated for all payoff streams $g$ of interest.

Sufficient conditions for $\UqX g(x)$ of a measurable stream
$g(X_t)$ to be finite are
\begin{eqnarray}\label{gpest}
|g(X_t)|&\le& C\exp(\sgp X_t),\quad X_t\ge 0,\\
\label{gmest}|g(X_t)|&\le& C\exp(\sgm X_t),\quad X_t\le 0,
\end{eqnarray}
where constant $C$ is independent of $X_t$, and $\sg^\pm$ satisfy
\begin{equation}\label{fin}
1-qM(\sg^\pm)>0.
\end{equation}
These conditions are necessary if $g$ is monotone on each
half-axis.

 If we confine ourselves to a wide and fairly flexible class of
probability densities given by exponential polynomials on each half-axis, then for a general $g$, the calculation
of $\UqX g(X)$ reduces to simple integration procedures.  Consider first the case when the transition density is
of the form
\begin{equation}\label{dens11} p(x)=\cp\lp e^{-\lp x}{\bf 1}_{[0,
+\infty)}(x)+\cm (-\lm)e^{-\lm x}{\bf
1}_{(-\infty,0]}(x),\end{equation}
  where $\cp, \cm>0$, and $\lm<0<\lp$.
   If we want to have a
  continuous $p$, we must require that $\cp\lp+\cm\lm=0$, and then the normalization requirement $M(0)=1$
  leads to $\cp=\lm/(\lm-\lp), \cm=\lp/(\lp-\lm)$. We have constructed a two-parameter family of
  probability densities.The moment
  generating function is
  \[
  M(z)=\frac{\cp\lp}{\lp-z}+\frac{\cm\lm}{\lm-z}=\frac{-\lm\lp}{\lp-\lm}\left[\frac{1}{\lp-z}-\frac{1}{\lm-z}\right].
  \]
  It is easily seen that $1-qM(z)$ has two real roots: $\bem\in
  (\lm, 0)$, and $\bep\in (0, \lp)$, which are the roots
  of the quadratic equation
  \begin{equation}\label{cheq11}
  z^2-(\lp+\lm)z+(1-q)\lp\lm=0.
  \end{equation}
  We find
  \begin{equation}\label{roots11}
\be^\pm=0.5\cdot\left(\lp+\lm \pm\sqrt{(\lp+\lm)^2-4(1-q)\lm\lp}\right),
\end{equation}
and represent
   $1/(1-qM(z))$ in the form
  \begin{equation}\label{res11}
\frac{1}{1-qM(z)}=\frac{\ap}{\bep-z}+\frac{\am}{\bem-z},
\end{equation}
where $a^\pm=(\be^\pm-z)/(1-qM(z))\vert_{z=\be^\pm}=1/qM'(\be^\pm)$.
Set $g(x)=e^{z x}$. Using \eq{dens11} and \eq{res11}, it is straightforward to check that the RHS in \eq{epv4} is
equal to the RHS in the equation below
\begin{equation}\label{epv11}
\UqX g(x)=\ap\int_0^\infty e^{-\bep y}g(x+y)dy-\am\int^0_{-\infty} e^{-\bem y}g(x+y)dy.
\end{equation}
By linearity, \eq{epv11} holds for linear combinations of
exponents, and more generally, for wide classes of functions which
can be represented as integrals of exponents $e^{z x}$ w.r.t. $z$.
One can also use more than one exponential on each axis, and
obtain more elaborate probability densities (see  Appendix B for
details).


\subsection{Payoff stream that starts to accrue at a random time}
As an example, consider the problem of timing an investment of a fixed size (see Subsection 2.4.1). We need to
compute  the value
given by \eq{whp}, where
\[
g(X_t)=qGE_t[e^{X_{t+1}}]-(1-q)C=qM(1)Ge^{X_t}-(1-q)C. \]
 The operator $(1-q)\UqX$ acts on an exponential function $e^{zx}$ as the multiplication operator by
 the number $(1-q)/(1-qM(z))$:
 \[
 (1-q)\UqX e^{zx}= \frac{1-q}{1-qM(z)}e^{zx}.
 \]
 Similarly, the
  $(1-q)\UqbX$ and $(1-q)\UquX$ act on an
 exponential function $e^{z x}$ as multiplication operators
 by certain numbers, call them $\kapq(z)$ and $\kamq(z)$:
\begin{equation}\label{actkapm}
(1-q)\UqbX e^{z x}=\kapq(z)e^{z x},\quad (1-q)\UquX e^{z
x}=\kamq(z)e^{z x}.
\end{equation}
These numbers are
 \[
\kapq(z)=(1-q)\left.(\UqbX e^{zx})\right|_{x=0}
 =(1-q)E\left[\sum_{t=0}^\infty q^t g(\barX_t)\ |\ X_0=0\right],
 \]
 and
\[
 \kamq(z)=(1-q)\left.(\UquX e^{zx})\right|_{x=0}=
  (1-q)E\left[\sum_{t=0}^\infty q^t g(\uX_t)\ |\ X_0=0\right],
\]
 respectively. The Wiener-Hopf factorization formula (see, e.g.,
 \cite{Sp}) states that
 \begin{equation}\label{wh0}
 \frac{1-q}{1-qM(z)}=\kapq(z)\kamq(z).
 \end{equation}
 It follows that \eq{wh} holds: $ \UqX
 =(1-q)\UqbX \UquX$. In other words, we can
 factorize the EPV--operator $\UqX$ in terms of the EPV--operators
 $\UqbX$ and $\UquX$.
Indeed, if we apply both sides of \eq{wh} to $e^{zx}$, we
 obtain identity on the strength of \eq{wh0}. By linearity,
 \eq{wh} extends to linear combinations of exponential functions, and more
 generally, to wide classes of functions which can be represented
 as integrals of exponentials w.r.t. parameter $z$.

Under an additional very weak regularity condition on the density $p$ (piece-wise continuous $p$ are allowed), it
is proved in \cite{BL2} that
 if $g$ is measurable and satisfies the growth condition
 \eq{gpest}, then for $\tau^+=\tau^+_h$, \eq{whp} holds.
 It is possible to prove \eq{whp} without additional conditions on $p$.

If the probability density $p$ is given by exponential polynomials
 on the positive and negative half-axis, then all the functions in \eq{wh0} are
 rational functions. The calculation of
 the factors $\ka^\pm_q(z)$ reduces to the calculation of roots of
 the numerator of the rational function $1-qM(z)$, that is, of
 a polynomial. To obtain $\kapq(z)$,
 one collects all the factors in the numerator and denominator of the rational function $(1-q)/(1-qM(z))$,
 which do not vanish in the half-plane $\Re z\ge 0$
 (the real part of $z$ is non-negative),
 and $\kamq(z)$ contains the factors which do not vanish in the half-plane
 $\Re z\le 0$ (see, e.g., \cite{Kemp} and \cite{BL2}. In many cases (including the examples
 above), all the roots are real. Now the same argument as at the end of Section 2
 shows that the action of operators $\UqbX$
 and $\UquX$ is given by simple integral operators. As the result, we obtain an effective procedure for the calculation of  $\UqbX g(x)$ and $\UquX g(x)$. In this
section, we consider the  case of the probability density
\eq{dens11}; for the general case, see Appendix B.

First, we calculate the the moment-generating function $M(z)$,
find the roots of the equation $1-qM(z)=0$ (see \eq{roots11}), and
represent $\kapq(z)$ and $\kamq(z)$ as the sums of constants and
simple fractions:
\begin{equation}\label{kapq11}
\kapq(z)=\frac{(\lp-z)\bep}{\lp(\bep-z)}=\frac{\bep}{\lp}+\frac{\bep(\lp-\bep)}{\lp(\bep-z)},
\end{equation}
and
\begin{equation}\label{kamq11}
\kamq(z)=\frac{(\lm-z)\bem}{\lp(\bem-z)}=\frac{\bem}{\lm}+\frac{\bem(\lm-\bem)}{\lm(\bem-z)}.
\end{equation}
Using \eq{actkapm}, \eq{kapq11}, \eq{kamq11} and the same considerations as in the derivation of \eq{epv11}, we
obtain for a continuous $g$ satisfying \eq{gpest}
\begin{equation}\label{actkapp11}
(1-q)\UqbX g(x)=\frac{\bep}{\lp}g(x)+\frac{\bep(\lp-\bep)}{\lp}
\int_0^{+\infty} e^{-\bep y}g(x+y)dy,
\end{equation}
and for a continuous $g$ satisfying \eq{gmest}, we calculate
\begin{equation}\label{actkamm11}
(1-q)\UquX g(x)=\frac{\bem}{\lm}g(x)-\frac{\bem(\lm-\bem)}{\lm}
\int_{-\infty}^0 e^{-\bem y}g(x+y)dy.
\end{equation}
Under condition \eq{fin}, the roots $\be^\pm$ are outside $[\sgm, \sgp]$, therefore \eq{gpest} and \eq{gmest}
ensure the convergence of the integrals in \eq{actkapp11} and \eq{actkamm11}.

Now we return to the investment problem of Subsection 2.4.  From
\eq{actkapm}, we find
\[
\UquX g(x)=(1-q)^{-1}\kamq(1)qM(1)Ge^x-C.
\]
Set \[ u(x)=\bfo_{[h, +\infty)}(x)  (\UquX g)(x)= \bfo_{[h,
+\infty)}(x)[(1-q)^{-1}\kamq(1)qM(1)Ge^x-C], \] and calculate the
value $V^+(x;h)=E^x[q^{\tau^+} \UqX g(X_{\tau^+})]$ of the
investment project at time 0 (provided $h$ is chosen as the
investment threshold).
 Using \eq{whp} and \eq{actkapp11},
we find for $x<h$:
\begin{eqnarray}\nonumber
V^+(x;h)&=&(1-q)(\UqbX u)(x)\\\nonumber
 &=&\frac{\bep(\lp-\bep)}{\lp}\int_{x+y>h}e^{-\bep y}
  [(1-q)^{-1}\kamq(1)qM(1)Ge^{x+y}-C]dy\\\nonumber
 &=&\frac{\bep(\lp-\bep)}{\lp}\left[\frac{\kamq(1)qM(1)G}{1-q}\int_{h-x}^{+\infty}e^{(1-\bep)y+x}dy
  -\int_{h-x}^{+\infty} C e^{-\bep y}dy\right]\\\label{val1}
 &=&\frac{\bep(\lp-\bep)}{\lp} e^{\bep
 (x-h)}\left[\frac{e^h\kamq(1)qM(1)G}{(1-q)(\bep-1)}-\frac{C}{\bep}\right].
 \end{eqnarray}

\subsection{Timing an investment} In the case of the stream $g(x)=e^x qM(1)G-(1-q)C$,
 \eq{thrp1} assumes the form
\begin{equation}\label{thr2} (1-q)^{-1}e^x\kamq(1)qM(1)G-C=0,
\end{equation}
and the investment threshold is
\begin{equation}\label{thrmm}
e^{h^\ast}=\frac{(1-q)C}{\kamq(1)qM(1)G}.
\end{equation}
After $h^\ast$ is found, the EPV of the project can be calculated
with the help of \eq{whp}. In particular, if the probability
density is given by \eq{dens11},  we can use \eq{val1} with
$h=h^*$,
 and  substituting \eq{thrmm}, find
\begin{equation}\label{epvinv2}
 V^+(x;h^\ast)=C\frac{(\lp-\bep) e^{\be^+
 (x-h^\ast)}}{(\bep-1)\lp}, \quad x<h^*.
 \end{equation}

\subsection{Timing an exit}
Consider a special case when the profit function in Subsection 2.4.2 is given as  $f(X_t)=Ge^{X_t}$. Then $g(x)$
is a linear combination of exponentials, so that we can calculate $\UqbX g(x)$ using \eq{actkapm}:
\[
\UqbX g(x)=q^2C-(1-q)^{-1}\kapq(1)qM(1)Ge^x,
\]
and find the optimal disinvestment threshold from \eq{exit1}:
\begin{equation}\label{thrpp1}
e^{h_\ast}=\frac{q^2(1-q)C}{\kapq(1)qM(1)G}.
\end{equation}
After $h_\ast$ is found from \eq{thrpp1}, we can calculate the
expected present value of the gains from scrapping
the production unit at time 0, 
$V^-(x;h_*)$, using \eq{whm} and \eq{thrpp1}. Similarly to \eq{epvinv2}, for 
$x>h_*$,
\[V^-(x; h_*)=Cq^2\frac{(\lm-\bem)e^{\be^-
 (x-h_\ast)}}{(1-\bem)\lm}.\]
 Finally, the value of the firm is the sum of the
EPV of the perpetual stream of profits plus the option value of scrapping: for $x>h_\ast$,
\[
V(x)=\frac{e^xG}{1-qM(1)}+Cq^2\frac{(\lm-\bem)e^{\be^-
 (x-h_\ast)}}{(1-\bem)\lm}.
\]

\section{Incremental capital expansion}
In this Section, we assume that the production function depends
only on capital, and
 that $G(K)$  is
differentiable, concave, and satisfies the Inada conditions. A
similar situation was considered in \cite{Ab} for a two-period
model of partially reversible investment, in \cite{DP} for the
geometric Brownian motion model, and in \cite{BO} for L\'evy
processes. At each time period $t$, the firm receives
$e^{X_t}G(K_t)$ from the sales of its product, and, should it
decide to increase the capital stock, suffers the installation
cost $C\cdot(K_{t+1}-K_t)$. The firm's objective is to chose the
optimal investment strategy $\cK=\{K_{t+1}(K_t, X_t)\}_{t\ge 1},
K_0=K, X_0=x$, which maximizes the NPV of the firm:
\begin{equation}\label{vf1}
V(K, x)= \sup_{\cK}E^x\left[\sum_{t\ge 0} q^{t}(e^{X_t}G(K_t)-
 C(K_{t+1}-K_t))\right].
 \end{equation}
Here we treat the current log price $x$ and capital stock $K$ as state variables, and $\cK$ as a sequence of
control variables. Due to irreversibility of investment, $K_{t+1}\ge K_t, \ \forall t$.

To ensure that firm's value \eq{vf1} is bounded, we impose a resource constraint: there exists $\bar{K}<\infty$,
such that $K_t\le\bar{K}, \ \forall t$.
The resource constraint, condition \eq{bd2}, and properties of the
production function ensure that the value function \eq{vf1} is
well defined.

Formally, the manager has to choose both the timing and the size
of the capital expansion. However, it is well-known (see, for
example, \cite{DP}) that for each level of the capital stock, it
is only necessary to decide when to invest. The manager's problem
is equivalent to finding the boundary (the investment threshold),
$h(K;C)$, between two regions in the state variable space $(K,x)$:
inaction and action ones. For all pairs $(K, x)$ belonging to the
inaction region, it is optimal to keep the capital stock
unchanged. In the action region, investment becomes optimal. To
derive the equation for the investment boundary, suppose first
that every new investment can be made in chunks of capital,
$\Delta K$, only\footnote{The authors are indebted for this
simplifying trick to Mike Harrison; our initial proof was more
involved.}. In this case, the firm has to suffer the cost $C\Delta
K$, and the EPV of the revenue gain due to this investment can be
represented in the form of the EPV of the stream
$g(X_t)=qM(1)(G(K+\Delta K)- G(K))e^{X_t}-(1-q)C\Delta K$. On the
strength of the result of Subsection 3.3, the optimal exercise
boundary is determined from the equation $\UquX g(h)=0$, which can
be written as
\begin{equation}\label{marsh1}
(1-q)^{-1}qM(1)(G(K+\Delta K)- G(K))\kamq(1)e^h=C\Delta K.
\end{equation}
 Dividing by $\Delta K$ in \eq{marsh1}
and passing to the limit, we obtain the equation for the optimal
threshold, $h^\ast=h^\ast(K)$:
\begin{equation}\label{marsh2}
qM(1)\kamq(1)G'(K)e^h=C(1-q). \end{equation} Equivalently, the optimal exercise price is
\begin{equation}\label{marsh3}
e^{h^\ast}=e^{h^\ast(K)}=\frac{C(1-q)}{qM(1)\kamq(1)G'(K)}.
\end{equation}
The rigorous justification of this limiting argument can be made exactly as in the continuous time model in
\cite{BO}. Let $h=h(K; \Delta)$ be the solution to \eq{marsh1}. Then the option value associated with the increase
of the capital by $\Delta K$, at the price level $e^x$, is
\[
(1-q)\UqbX \bfo_{[h, +\infty)}(x)
\left(\frac{qM(1)}{1-q}(G(K+\Delta K)- G(K))\kamq(1)e^x-C\Delta
K\right).
\]
As $\Delta K\to 0$, we have $h=h(K; \Delta)\to h^\ast(K)$;
therefore, dividing by $\Delta K$ and passing to the limit, we
obtain the formula for the derivative of the option value of
future investment opportunities w.r.t. $K$:
\begin{equation}\label{valmarsh1}
V^{\mathrm{opt}}_K(K, x)=(1-q)\UqbX \bfo_{[h^\ast, +\infty)}(x) \left(\frac{qM(1)G'(K)}{1-q}\kamq(1)e^x-C\right).
\end{equation}
Substituting $C$ from \eq{marsh3} into \eq{valmarsh1} and using
the definition of $\kamq(1)$, we obtain \[ V^{\mathrm{opt}}_K(K,
x)=(1-q)E\left[\sum_{t=1}^\infty
q^tG'(K)e^{\uX_t}\,\vert\,X_0=0\right]E^x\left[\sum_{t=0}^\infty
q^t(e^{\barX_t}-e^{h^*})_+\right].\] The last formula  factors out
the contributions of the infimum and supremum price processes to
the marginal option value of capital. The first expectation on the
RHS decreases if the probability of downward jumps in prices
increases, and the second expectation increases if the probability
of positive jumps in prices increases. Hence the marginal option
value of capital increases in downward uncertainty and decreases
in upward uncertainty. The overall effect of uncertainty is
ambiguous.

Now the marginal, or shadow, value of capital is given by
\begin{equation}\label{ww01}
V_K(K, x)= \frac{e^xG'(K)}{1-qM(1)} -V^{{\mathrm{ opt}}}_K(K, x).
\end{equation}
Equation \eq{ww01} expresses the marginal value of capital as the
difference of  two components. The first one is the expected
present value of the marginal returns to capital, given that the
capital stock remains constant at the level $K$ in the future. The
second component is the marginal option value of the future
investment opportunities. This value is subtracted because
investing extinguishes the option. A similar result was obtained
in \cite{Ab} for a two-period model of partially reversible
investment.

 Exactly the same calculations which lead to \eq{epvinv2} allow us to derive from \eq{valmarsh1}
\begin{equation}\label{valmarsh2}
V^{\mathrm{opt}}_K(K, x)=C\frac{(\lp-\bep) e^{\be^+
 (x-h^\ast)}}{(\bep-1)\lp}.
\end{equation}
Substituting \eq{marsh3} into the last equation, we derive
\begin{equation}\label{valmarsh3} V^{\mathrm{opt}}_K(K, x)=C \frac{\lp-\bep
}{(\bep-1)\lp}\left(\frac{\kamq(1)qM(1)}{(1-q)C}\right)^{\bep}e^{\bep x}G'(K)^{\bep}.
\end{equation}
Integrating \eq{valmarsh3} w.r.t. capital, we obtain
\[ V^{\mathrm{opt}}(K, x)=C\frac{\lp-\bep
}{(\bep-1)\lp}\left(\frac{\kamq(1)qM(1)}{(1-q)C}\right)^{\bep}e^{\bep x}\int_{K}^{\bar K}G'(k)^{\bep}dk.
\] The value of the firm  is
$V(K,x)=e^xG(K)/(1-qM(1))+V^{\mathrm{opt}}(K,x)$. Suppose that the
capital stock available for investment, $\bar{K}$, is very large,
and  function $G'(K)^{\bep}$ is integrable on $[1, +\infty)$. Then
we can obtain a simpler formula by replacing the upper limit $\bar
K$ with $+\infty$. In the case of Cobb-Douglas production function
$G(K)=dK^\theta$, we have $G'(K)=d\theta K^{\theta-1}$, therefore
the integrals converge iff $(\theta-1)\bep<-1$, or, equivalently,
$\bep>1/(1-\theta)$.
If this condition is satisfied, the value of the firm for
$x<h^*(K)$ is
\begin{eqnarray*}
V(K,x)& =&\frac{e^xdK^\theta}{1-qM(1)}\\
&&+\frac{\lp-\bep
}{(\bep-1)\lp}\left(\frac{\kamq(1)qM(1)}{(1-q)}\right)^{\bep}
\frac{C^{1-\bep}e^{\bep x}K^{1-\bep
(1-\theta)}}{(d\theta)^{\bep}(\bep(1-\theta)-1)}.
\end{eqnarray*}
One of the standing assumptions in corporate finance is that the value of a firm follows a (geometric) Brownian
motion.  Our solution clearly shows that the firm's value is not described even by a Markovian process.

Notice that the proof of \eq{marsh2} in \cite{BO} was based on the reduction to the case of the perpetual American
call, and therefore the generalization for more general dependence on the stochastic factor was not possible. Here
the result holds for any continuous increasing revenue flow $R(K, x)$, and  the formula for the optimal investment
threshold obtains in the form:
\begin{equation}\label{marsh4}
\UquX qT R_K(K, h)=C,
\end{equation}
where the EPV--operator $\UquX$ and transition operator $T$ act
w.r.t. the second argument (we have $qTR_K$ instead of $R_K$
because the revenues will start to accrue the period after the
investment is made). For instance, if the firm faces the
operational cost $a+bK e^{X_t/2}$, then the revenue flow is
$R(K_t, X_t)=e^{X_t}G(K_t)-a-bKe^{X_t/2}$, and instead of
\eq{marsh3}, we now have
\begin{equation}\label{marsh5}
(1-q)^{-1}[qM(1)\kamq(1)G'(K)e^{h}-bqM(1/2)\kamq(1/2)e^{h/2}]-C=0.
\end{equation}
The function on the LHS in \eq{marsh5} changes sign only once, and therefore the solution to equation \eq{marsh5}
gives the optimal investment threshold.

\appendix

\section{Proof of optimality of $h^\ast$ and $h_*$}

 On the strength of  Lemma on p.1364 in \cite{DLT}, to prove the optimality  of the solution $V(h^\ast;
x)$, it suffices to check the following two conditions:
\begin{equation}\label{opt1}
V(h^\ast; x)\ge \max\{G(x), 0\},\quad \forall\ x,
\end{equation}
 and
\begin{equation}\label{opt2}
(I-qT)V(h^\ast; x)\ge 0,\quad \forall\ x.
\end{equation}
By our choice of $h^*$, $w(x)\ge 0$ for all $x\ge h^\ast$, therefore $V(h^*; x)=(1-q)\UqbX \bfo_{[h^*,
+\infty)}w(x)\ge 0$ for all $x$. Using \eq{whp}, we represent $V(h^*; x)$ in the form
\begin{eqnarray*}
V(h^*; x) &=&(1-q)\UqbX \UquX g(x)-(1-q)(\UqbX {\bf 1}_{(-\infty,
h^*)}w)(x)\\
&=& \UqX g(x) - v(h^*; x)=G(x)-v(h^*; x),
\end{eqnarray*}
where $v(h^*; x)=(1-q)(\UqbX {\bf 1}_{(-\infty, h^*)}w)(x)$. Due
to the choice of $h^*$, we have $w(x)<0$ $\forall\,x<h^*$, so that
$v(h^\ast; x)\le 0$ for all $x$, and hence $V(h^\ast; x)\ge G(x)$
for all $x$. We conclude that \eq{opt1} holds.

  Under
a very weak regularity assumption on $p$,   it is proved in \cite{BL2} that for any $h$,
\begin{equation}\label{bell}
V(h; x)=qTV(h; x),\quad\ x<h.
\end{equation}
Using more elaborate arguments as in \cite{BL4} and \cite{BL2} in
continuous time, one can prove that \eq{bell} holds if the
probability density $p$ exists. Thus, \eq{opt2} holds on
$(-\infty, h^\ast)$, and
 it remains to verify \eq{opt2} for $x\ge h^\ast$. Introduce $
W(x)=(I-qT)V(h^\ast; x)$. Using \eq{trans} and \eq{wh}, we obtain
\begin{eqnarray}\nonumber
W(x)&=&(I-qT)(1-q)\UqbX \bfo_{[h^\ast, +\infty)}\UquX g(x)\\\label{w0}
 &=&(\UquX)^{-1}\bfo_{[h^\ast, +\infty)}\UquX g(x),
 \end{eqnarray}
 and also
\begin{eqnarray*}
W(x)&=&(I-qT)(1-q)\UqbX \bfo_{[h^\ast, +\infty)}\UquX g(x)\\
&=&(I-qT)(1-q)\UqbX \UquX g(x)-
(I-qT)(1-q)\UqbX \bfo_{(-\infty, h^\ast)}\UquX g(x)\\
&=&g(x) +(-I+qT)(1-q)\UqbX \bfo_{(-\infty, h^\ast)}\UquX g(x).
\end{eqnarray*}
But if a function $u$ vanishes outside $(-\infty, h)$, then
\[
\UqbX u(x)=E\left[\sum_{t=0}^\infty q^t
u(x+\barX_t)\right]=0,\quad\forall\ x\ge h,
\]
as well. Therefore, for $x\ge h^\ast$,
\begin{equation} \label{w1}
W(x)=g(x)+q(1-q)T\UqbX \bfo_{(-\infty, h^\ast)}\UquX g(x).
\end{equation}
By our standing assumption, $g$ is non-decreasing, hence from \eq{w1}, $W(x)$ is non-decreasing on $[h^\ast,
+\infty)$. We apply $\UquX$ to \eq{w0} and obtain
\begin{equation}\label{w2} \bfo_{[h^\ast,
+\infty)}(x)\UquX g(x)=\UquX W(x)=E\left[\sum_{t=0}^\infty q^t W(x+\uX_t)\right].
\end{equation}
 Suppose that
$W(h^\ast)<0$. Then there exists $h_1>h^\ast$ such that $W(x)\le
0$ for all $x\in (h^\ast, h_1)$. It follows that for the same $x$,
the RHS in \eq{w2} is non-positive. But for these $x$, the LHS is
positive by the very definition of $h^\ast$. Hence, our assumption
$W(h^\ast)<0$ is false, and since $W$ is non-decreasing on
$[h^\ast, +\infty)$, the condition \eq{opt2} follows, and the
proof of optimality is finished.

The verification of the optimality conditions \eq{opt1}--\eq{opt2}
for the exit problem is quite similar to the proof above: just
replace $\barX, \uX$, $\UqbX, \UquX$ and $[h^\ast, +\infty)$ with
$\uX, \barX, \UqbX, \UquX$ and $(-\infty, h_\ast]$, respectively.

\section{Transition densities given by exponential polynomials}
\subsection{The case of three exponentials}
In this Subsection, we demonstrate how to obtain the transition density of a desired shape.  The density
\eq{dens11} has a kink (and maximum) at the origin. If we want to have a smooth $p$ (and allow for the maximum to
be not at the origin), we need to use more than two exponential functions. Suppose that we want to model a density
which has the maximum on the positive half-axis. Then we use one exponential on the negative half-axis, and two on
the positive one:
\begin{equation}\label{dens12} (\cpo\lpo
e^{-\lpo x}-\cpt\lpt e^{-\lpt x}){\bf 1}_{[0, +\infty)}(x)-\cm \lm
e^{-\lm x}{\bf 1}_{(-\infty,0]}(x),\end{equation} where $\cpo,
\cpt$ and $\cm$ are positive, and $\lm<0<\lpo<\lpt$. Later in this
Subsection, we show that
 for any choice of
$\lm<0<\lpo<\lpt$, equation \eq{dens12} with $\cm$, $\cpo$, $\cpt$
given by simple formulas \eq{solc} defines a probability density,
which has the maximum on the positive half-axis. See Figure 1 for
an example. Similarly, one can construct a 3-parameter family of
probability densities  which have the maximum on the negative
half-axis. Should one wish to have a smooth probability density
which has the maximum at the origin, one must use two exponential
functions on each half-line or exponential polynomials of the form
$(ax+b)e^{\ga x}$.

The moment generating function of the probability density
\eq{dens12} is
\[
M(z)=\frac{\cpo\lpo}{\lpo-z}-\frac{\cpt\lpt}{\lpt-z}+\frac{\cm\lm}{\lm-z}.
\]
At the end of this Subsection we will show that the {\em
fundamental rational function} $1-qM(z)=0$ has 3 roots, all of
which real. Call these roots $\bem$, $\bepo$ and $\bept$. We have
\begin{equation}\label{roots12}
\lm<\bem<0<\bepo<\lpo<\lpt<\bept.
\end{equation}
Clearly, $1/(1-qM(z))$ can be represented in the form
  \begin{equation}\label{res12}
\frac{1}{1-qM(z)}=\frac{\apo}{\bepo-z}+\frac{\apt}{\bept-z}+\frac{\am}{\bem-z},
\end{equation}
where $a_{+,j}=1/(qM'(\be^+_j)),\ j=1,2,$ and $\am=1/(qM'(\bem))$, and therefore,
\begin{eqnarray}\label{epv12}
\UqX g(x)&=&\apo\int_0^\infty e^{-\bepo y}g(x+y)dy\\\nonumber &+&\apt\int_0^\infty e^{-\bept
y}g(x+y)dy-\am\int^0_{-\infty} e^{-\bem y}g(x+y)dy.
\end{eqnarray}
Similarly, we can consider a probability density given by linear combinations of two or more exponents on each of
the half-axis. If we use two exponentials for each, we have two roots $\be^\pm_j$, $j=1,2,$ of the
``characteristic equation" $1-qM(z)=0$ on each half-axis, and \eq{roots12}--\eq{epv12} change in the
straightforward manner. One can also use more than two exponentials on each axis, and obtain more elaborate
probability densities.

Now we show that any choice $\lm<0<\lpo<\lpt$ defines a
probability density. Three conditions: $\INT p(x)dx=1$, $p$ is
continuous at 0, and  $p$ is smooth at 0, give a linear system of
three equations
\begin{eqnarray}\nonumber
\cpo-\cpt+\cm&=&1,\\\label{eqc}
 \cpo\lpo-\cpt\lpt+\cm\lm&=&0,\\\nonumber
 \cpo(\lpo)^2-\cpt(\lpt)^2+\cm(\lm)^2&=&0.
 \end{eqnarray}
 Using Cramer's rule, it is easy to find a unique solution
 $(\cpo, \cpt, \cm)$ to \eq{eqc} for any $\lm<\lpo<\lpt$:
 \begin{eqnarray}\nonumber
 \cpo&=&\frac{-\lm\lpt}{(\lpt-\lpo)(\lpo-\lm)},\\\label{solc}
 \cpt&=&\frac{-\lm\lpo}{(\lpt-\lpo)(\lpt-\lm)},\\\nonumber
 \cm&=& \frac{\lpo\lpt}{(\lpo-\lm)(\lpt-\lm)}.
  \end{eqnarray}
It is easily seen that $\cpo, \cpt$ and $\cm$ are positive, and
that $p$ is positive as well.

The roots of $1-qM(z)$ are found as follows. Clearly, $1-qM(z)$
has 3 roots at most. As $z\to\lm+0$, $1-qM(z)\to-\infty$,  and the
same holds as $z\to\lpo-0$, and as $z\to \lpt+0$. Under condition
$q\in (0,1)$, $1-qM(0)=1-q>0$, and $1-qM(+\infty)=1>0$. Hence, on
each of the intervals $(\lm, 0)$ $(0, \lpo)$, and $(\lpt,
+\infty)$, $1-qM(z)$ changes sign. Therefore, on each of these
three intervals, there is exactly one root, which we have called
$\bem$, $\bepo$ and $\bept$, respectively.

\subsection{General scheme for the computation of $\UqbX$ and
$\UquX$} {\bf Step 1.} Calculate the moment-generating function
$M(z)$, and consider the rational function $1-qM(z)$. Find the
roots of the denominator, $\la^\pm_j,$ and the numerator,
$\be^\pm_j$, with their multiplicities (sign ``+" for the roots on
the positive axis, sign ``-" for the ones on the negative axis).

\noindent {\bf Step 2.} Define
\begin{equation}\label{kap}
\kapq(z)=\prod_{j}\frac{\la^+_j-z}{\la^+_j}\prod_{k}\frac{\be^+_k}{\be^+_k-z},
\end{equation}
\begin{equation}\label{kam}
\kamq(z)=\prod_{j}\frac{\la^-_j-z}{\la^-_j}\prod_{k}\frac{\be^-_k}{\be^-_k-z}.
\end{equation}

\noindent{\bf Step 3.} If all the roots $\be^\pm_k$ are simple,
we represent $\kapq(z)$ and $\kamq(z)$ in the form
\begin{equation}\label{kapm}
\kapq(z)=\kapq(\infty)+\sum_k \frac{a^+_k}{\be^+_k-z},\quad
\kamq(z)=\kamq(\infty)-\sum_k \frac{a^-_k}{\be^-_k-z},
\end{equation}
where
\[
a^+_k=\prod_{j}\frac{\la^+_j-\be^+_k}{\la^+_j}\be^+_k\prod_{l\neq
k}\frac{\be^+_l}{\be^+_l-\be^+_k},\quad
a^-_k=-\prod_{j}\frac{\la^-_j-\be^-_k}{\la^-_j}\be^-_k\prod_{l\neq
k}\frac{\be^-_l}{\be^-_l-\be^-_k}.
\]
The case of multiple roots can be treated similarly.

\noindent{\bf Step 4.} For a continuous $g$ satisfying \eq{gpest},
we can calculate
\begin{equation}\label{actkapp}
\UqbX g(x)=\kapq(\infty)g(x)+\sum_k a^+_k \int_0^{+\infty}
e^{-\be^+_k y}g(x+y)dy,
\end{equation}
and for a continuous $g$ satisfying \eq{gmest}, we can find
\begin{equation}\label{actkamm}
\UquX g(x)=\kamq(\infty)g(x)+\sum_k a^-_k \int_{-\infty}^0
e^{-\be^-_k y}g(x+y)dy.
\end{equation}
Under condition \eq{fin}, all the roots $\be^\pm_k$ are outside $[\sgm, \sgp]$, therefore \eq{gpest} and
\eq{gmest} ensure the convergence of the integrals in \eq{actkapp} and \eq{actkamm}.

\subsection{Value of investment project} By assuming that the probability
density is given by exponential polynomials on each half-axis and
using \eq{whp} and \eq{actkapp}, we find for $x<h$:
\begin{eqnarray}\nonumber
V^+(x;h)&=&(1-q)(\UqbX u)\\\nonumber
 &=&\sum_k a^+_k\left[\frac{\kamq(1)qM(1)G}{1-q}\int_{h-x}^{+\infty}e^{-\be^+_ky+x+y}dy
  -C
 \int_{h-x}^{+\infty}e^{-\be^+_ky}dy\right]\\\label{val1m}
 &=&\sum_k a^+_k e^{\be^+_k
 (x-h)}\left[\frac{e^h\kamq(1)qM(1)G}{(1-q)(\be^+_k-1)}-\frac{C}{\be^+_k}\right].
 \end{eqnarray}
 If the threshold is chosen optimally, then we use \eq{thrmm} and
 find
\begin{equation}\label{epvinv2m}
 V^+(x;h^\ast)=C\sum_k \frac{a^+_k e^{\be^+_k
 (x-h^\ast)}}{(\be^+_k-1)\be^+_k}.
 \end{equation}

\subsection{Value of scrapping} After $h_\ast$ is found from \eq{thrpp1}, we can calculate
the expected present value of the gains from scrapping of the firm
at time 0, $V^-(x;h_*)$,  using \eq{whm} and \eq{thrpp1}.
Similarly to \eq{epvinv2m}, for $x>h_\ast$,
\[V^-(x; h_*)=Cq^2\sum_k\frac{\am_k e^{\bem_k
 (x-h_\ast)}}{(1-\bem_k)\bem_k}.\]
 Finally, the value of the firm is the sum of the
EPV of the perpetual stream of profits plus the option value of scrapping: for $x>h_\ast$,
\[
V(x)=\frac{e^xG}{1-qM(1)}+Cq^2\sum_k\frac{\am_k e^{\bem_k
 (x-h_\ast)}}{(1-\bem_k)\bem_k}.
\]

\subsection{Capital expansion}
Exactly the same calculations which lead to \eq{epvinv2} allow us
to derive from \eq{valmarsh2}
\begin{equation}\label{valmarsh2m}
V^{\mathrm{opt}}_K(K,x)=C\sum_k \frac{a^+_k
}{(\be^+_k-1)\be^+_k}\left(\frac{\kamq(1)qM(1)}{(1-q)C}\right)^{\be^+_k}e^{\be^+_kx}G'(K)^{\be^+_k}.
\end{equation}
Integrating \eq{valmarsh2m} w.r.t. $K$, we obtain
\[
V^{\mathrm{opt}}(K,x)=C\sum_j \frac{a^+_j
}{(\be^+_j-1)\be^+_j}\left(\frac{\kamq(1)qM(1)}{(1-q)C}\right)^{\be^+_j}e^{\be^+_jx}\int_{K}^{\bar
K}G'(k)^{\be^+_j}dk.
\]
Suppose that the available capital stock, $\bar{K}$, is very
large, and the functions $G'(K)^{\be^+_k}$ are integrable on $[1,
+\infty)$. Then we can obtain a simpler formula by replacing the
upper limit $\bar K$ with $+\infty$. In the case of  Cobb-Douglas
production function $G(K)=dK^\theta$, we have $G'(K)=d\theta
K^{\theta-1}$, therefore the integrals converge iff
$(\theta-1)\be^+_k<-1$ for all $k$. Let $\be^+_1$ be the smallest
positive root, then the equivalent condition is
$\be^+_1>1/(1-\theta)$.
If this condition is satisfied, the option value of investment
opportunities is
\[
V^{\mathrm{opt}}(K,x)=\sum_k \frac{a^+_k
}{(\be^+_k-1)\be^+_k}\left(\frac{\kamq(1)qM(1)}{(1-q)}\right)^{\be^+_k}
\frac{C^{1-\be^+_k}e^{\be^+_kx}K^{1-\be^+_k(1-\theta)}}{(d\theta)^{\be^+_k}(\be^+_k(1-\theta)-1)}.
\]

\section{Expected waiting time}  Assume that the spot
log-price $x$ is less than $h^\ast$, and consider the waiting time
$R_x$ till the investment is made. If the transition density is
given by \eq{dens11}, then the expected waiting time is finite iff
$\lp+\lm<0$, or, equivalently,
\begin{equation}\label{c1} m\equiv
E[X_1]\equiv 1/\lm+1/\lp>0,
\end{equation}
and if \eq{c1} holds, then
\begin{equation}\label{rwait1}
E[R_x]=\frac{1}{m}(h^\ast-x+1/\lp).
\end{equation}
Condition \eq{c1} has a clear interpretation: the expected waiting time is finite iff the drift of the log-price,
$m$, is positive, and if it is positive, then \eq{rwait1} says that the expected waiting time is inversely
proportional to the drift. It is also proportional to the distance to the barrier {\em plus the positive term
which is independent of the distance}.
\vskip1.5cm

To obtain the formula \eq{rwait1} for the expected waiting time, we need to calculate the limit in \eq{wait0}.
Using \eq{actkapp11}, we obtain for $x<h^\ast$:
\begin{eqnarray*}
\UqbX \bfo_{[h^\ast, +\infty)}(x)&=&(1-q)^{-1}\frac{\bep(\lp-\bep)}{\lp}
\int_0^{+\infty} e^{-\bep y}\bfo_{[h^\ast, +\infty)}(x+y)dy\\
&=&\frac{ e^{-\bep
(h^\ast-x)}(\lp-\bep)}{(1-q)\lp}=(1-q)^{-1}e^{-\bep
(h^\ast-x)}(1-\bep/\lp).
\end{eqnarray*}
Hence, we have to calculate the limit  \[ \lim_{q\to 1-0} \frac{1-e^{-\bep(q) (h^\ast-x)}+e^{-\bep(q)
(h^\ast-x)}\bep(q)/\lp}{1-q}.
\]
Both terms are positive, and if $\bep(q)/(1-q)$ is unbounded as $q\to 1$, the limit is clearly infinite. From
\eq{roots11}, we find that this possibility does not realize iff $\lm+\lp<0$. If $\lm+\lp<0$,  we obtain
\[
\bep(q)=\frac{\lp\lm}{\lp+\lm}(1-q)+O((1-q)^2),
\]
and \eq{rwait1} follows.

\newpage

\begin{figure}
 \scalebox{0.7}
{\includegraphics{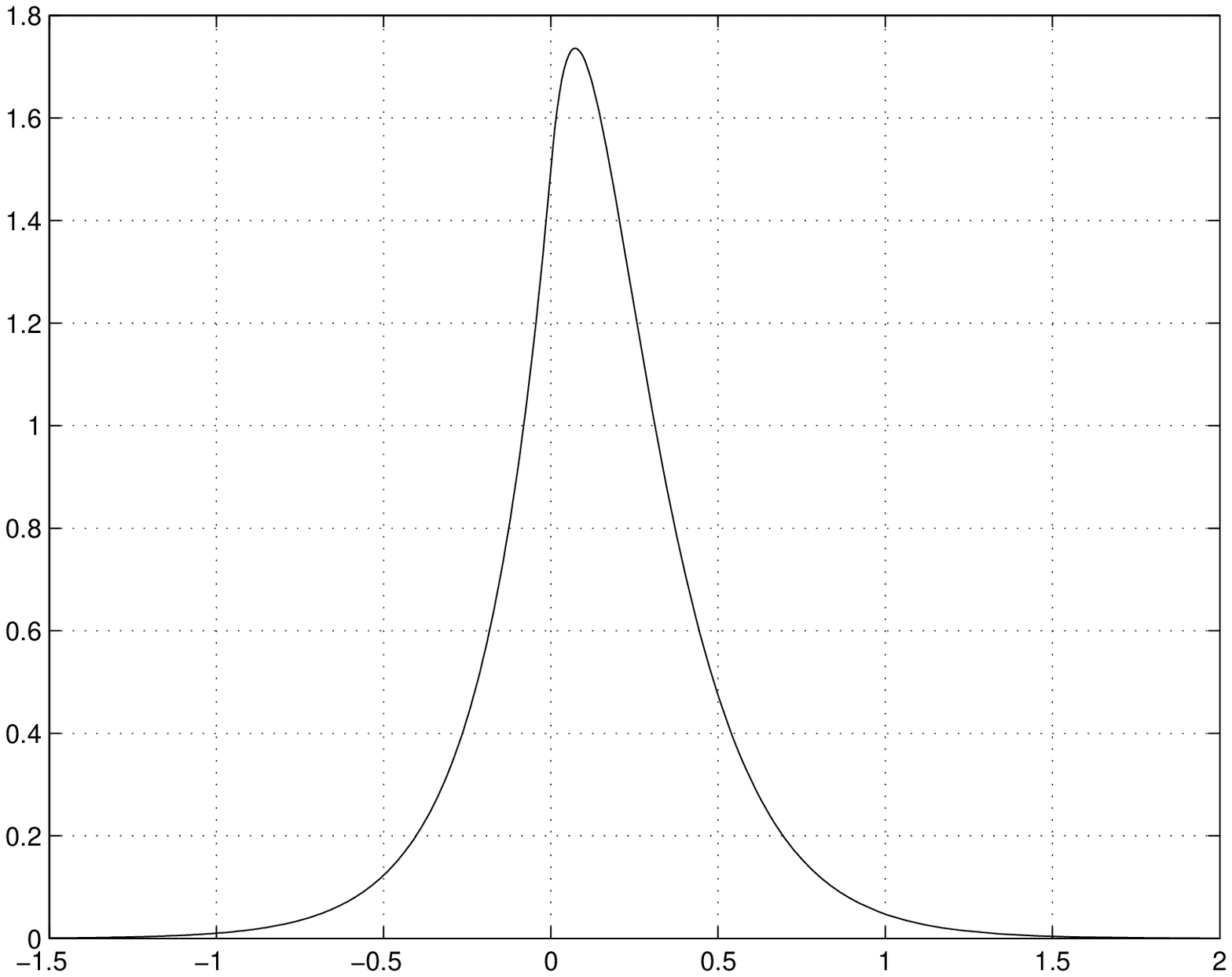}}
 \caption{Density \eq{dens12}. Parameters:
 $ \lm=-5, \lpo=5, \lpt=7.5$}
 \end{figure}

\end{document}